\documentclass[fleqn,usenatbib]{mnras}
\usepackage{newtxtext,newtxmath}
\usepackage[T1]{fontenc}
\DeclareRobustCommand{\VAN}[3]{#2}
\let\VANthebibliography\thebibliography
\def\thebibliography{\DeclareRobustCommand{\VAN}[3]{##3}\VANthebibliography}

\usepackage{graphicx}	
\usepackage{amsmath}	
\usepackage[flushleft]{threeparttable}
\usepackage{tabularx}

\newcommand\ergs{erg~s$^{-1}$}
\newcommand\ergcms{erg~cm$^{-2}$~s$^{-1}$}

\title[Single-frame rapid X-ray transients]{Searching for single-frame rapid X-ray transients detected with Chandra}

\author[Zhang \& Feng]{
Yijia Zhang$^{1}$
and Hua Feng$^{1,2}$\thanks{E-mail: hfeng@tsinghua.edu.cn (HF)}
\\
$^{1}$Department of Astronomy, Tsinghua University, Beijing 100084, China\\
$^{2}$Department of Engineering Physics, Tsinghua University, Beijing 100084, China\\
}

\date{Accepted XXX. Received YYY; in original form ZZZ}
\pubyear{2023}

\begin{document}
\label{firstpage}
\pagerange{\pageref{firstpage}--\pageref{lastpage}}
\maketitle

\begin{abstract}
We propose a new method to identify rapid X-ray transients observed with focusing telescopes. They could be statistically significant if three or more photons are detected with Chandra in a single CCD frame within a point-spread-function region out of quiescent background. In the Chandra archive, 11 such events are discovered from regions without point-like sources, after discrimination of cosmic rays and background flares and control of false positives. Among them, two are spatially coincident with extended objects in the Milky Way, one with the Small Magellanic Cloud, and another one with M31; the rest have no or a dim optical counterpart ($\gtrsim 20$~mag), and are not clustered on the Galactic plane. Possible physical origins of the rapid transients are discussed, including short gamma-ray bursts (GRBs), short-lived hypermassive neutron stars produced by merger of neutron stars, accreting compact objects in the quiescent state, magnetars, and stellar flares. According to the short GRB event rate density, we expect to have detected $2.3_{-0.6}^{+0.7}$ such events in the Chandra archive. This method would also allow us to reveal quiescent black holes with only a few photons. 
\end{abstract} 


\section{Introduction}

The X/gamma-ray universe is highly dynamic and exhibits a variety of transient phenomena over a wide range of energy and time scales.  Transients of short durations, i.e., those occurring on a timescale on the order of seconds or less, may represent violent energy release in a compact region.  For example, gamma-ray bursts (GRBs), which are associated with the collapse of massive stars \citep{Stanek2003} or merger of neutron stars \citep{Abbott2017}, can release an energy up to $10^{52}$~erg \citep{Cenko2010} over a duration typically of seconds, manifesting themselves the most luminous transient events in the universe \citep{Kumar2015}. Similar but less energetic rapid high energy transients include the soft gamma-ray repeaters associated with strongly magnetized neutron stars \citep{Mereghetti2008},  type-I thermal nuclear bursts on the surface of neutron stars \citep{Chakrabarty2003}, solar/stellar flares caused by magnetic activities on the surface of normal stars \citep{Benz2017}, and terrestrial gamma-ray flashes occurring in the atmosphere of the Earth \citep{Fishman1994}. 

Over the past decades, there have been many missions or instruments dedicated to the search of rapid X/gamma-ray transients \citep[e.g.,][]{Fishman1993,Boella1997,Kawai1999,Gehrels2004,Meegan2009,Wen2019,Li2020}. These instruments all have a wide field of view to increase the probability of catching a transient source from an unknown location.  Such a design, however, also increases the acceptance of background and consequently degrades the sensitivity. Therefore, all these instruments have a sensitivity much lower than that of focusing telescopes, and are only sensitive to bright transient sources like GRBs.  On the contrary, focusing telescopes such as the Chandra X-ray Observatory \citep{Weisskopf2000} have an unprecedented sensitivity, but their narrow field of view limits the chance of detecting a transient at a random location. 

Inspired by the detection of fast radio bursts \citep[for a review see][]{Zhang2020}, there may exist some high energy transients with quite low peak fluxes or fluences but appearing at a high occurrence rate.  This is a reasonable conjecture because luminosity functions of astrophysical objects generally have a power-law form with a negative power-law index. Then, they may have a better chance to be detected by high-sensitivity focusing telescopes despite the small field of view. Indeed, there have been reports of serendipitous discovery of transient X-ray sources with Chandra \citep[e.g.,][]{Irwin2016,Bauer2017,Xue2019,Yang2019,Quirola-Vasquez2022} during observations of other targets.  These events are not bright enough to trigger the wide-field monitors mentioned above.  

These transient sources were identified because they resulted in enough photons in the detector so that they appeared like ``sources'' in time-integrated images and were detected by regular source detection algorithms.  Then, their transient nature was determined by further timing analysis. However, a regular detection becomes impossible if a transient has a very short duration or low fluence, and consequently not enough photons to justify a significant source detection over a long exposure. For instance, a millisecond burst with a few photons stopped by the detector will be overwhelmed in the background if the observation has a long exposure (typically, tens of kiloseconds for Chandra observations). Such rapid but faint transients, however, could be significant in single-frame images of the CCD. In particular, Chandra has a superb angular resolution and thus extremely low background in the point-spread-function (PSF) region, such that a few photons during a single frame (several seconds) within the PSF region (a few pixels across) can be statistically significant given the background distribution.  The earlier search algorithms \citep[e.g.,][]{Yang2019} cannot reach such a low count threshold in order to compete the accumulated background over a long exposure. A new algorithm is needed to search for transients in a different parameter space. The high angular resolution of Chandra also allows for a precise determination of the transient location and identification of their multiwavelength counterparts, if any.  Random transients with low peak fluxes and short durations have not been systematically searched so far; they could be nearby events with low energetics/luminosities or distant ones produced by more violent physical processes.  

In this paper, we propose a new algorithm to search for rapid transients, those with a duration close to or less than the frame time, in the Chandra archive. We describe the algorithm for transient search in \S~\ref{sec:method}, including the discrimination of cosmic rays or background flares in \S~\ref{sec:contamination} and control of false positives in \S~\ref{sec:sig}. The candidate events and their possible optical counterparts are elaborated in \S~\ref{sec:result}. We discuss their physical natures in \S~\ref{sec:discuss}. 

\section{Data and search algorithm}
\label{sec:method}

\begin{table}
\footnotesize
\caption{Configurations of Chandra ACIS imaging observations used in our search.}
\label{tab:config}
\begin{threeparttable}
\begin{tabularx}{\columnwidth}{XXXX}
\hline\noalign{\smallskip}
$t_{\rm frm}$ & $n_{\rm frm}$ & FoV & $\log p_{\rm th}$  \\ 
(s) &  & (arcmin$^2$) & \\ 
\noalign{\smallskip}\hline\noalign{\smallskip}
\multicolumn{4}{c}{ACIS-I}\\ 
\noalign{\smallskip}\hline\noalign{\smallskip}
3.1 & 35292448 & 113.1 & -9.4\\ 
3.2 & 12052989 & 113.1 & -8.7\\ 
\noalign{\smallskip}\hline\noalign{\smallskip}
\multicolumn{4}{c}{ACIS-S2}\\ 
\noalign{\smallskip}\hline\noalign{\smallskip}
3.1 & 25879038 & 30.9 &  \\ 
3.2 & 17210557 & 30.9 &  \\ 
\noalign{\smallskip}\hline\noalign{\smallskip}
\multicolumn{4}{c}{ACIS-S3}\\ 
\noalign{\smallskip}\hline\noalign{\smallskip}
3.1 & 26183871 & 61.0 &  \\ 
3.2 & 17339909 & 61.0 & -9.1\\ 
\noalign{\smallskip}\hline\noalign{\smallskip}
\end{tabularx}
\begin{tablenotes}
\footnotesize
\item {\bf Note}. $t_{\rm frm}$ is the CCD frame time. $n_{\rm frm}$ is the total number of frames in that configuration in all observations. FoV is the effective field of view, determined by the CCD region inside a 6\arcmin-radius circle around the optical axis. $p_{\rm th}$ is the threshold $p$-value after the control of FDR.
\end{tablenotes}
\end{threeparttable}
\end{table}

We first make a rough estimate about how a single-frame transient can be detected. For example, the total background on the Advanced CCD Imaging Spectrometer (ACIS) S3 chip is roughly $1.9 \times 10^{-6}$ counts~s$^{-1}$~arcsec$^{-2}$ in the energy band of 0.5--7 keV\footnote{See the Chandra Proposers' Observatory Guide at \url{https://cxc.harvard.edu/proposer/POG/html/chap6.html}.}. Considering a PSF radius of 3\arcsec\ and a frame time of 3.2~s, the mean background counts are $1.7 \times 10^{-4}$ in a single frame within a PSF region, and the chance to see 3 or more counts within a PSF region at any location in the CCD (70.5~arcmin$^2$) is $7.6 \times 10^{-9}$ assuming Poisson distribution. This is the chance probability in a single frame of S3, and one must consider the number of trials that is the total number of frames (up to $10^7$) in which the same search is conducted. Therefore, finding at least 3 photons in a single frame within a PSF region out of pure background may lead to identification of transient events, if the false positives can be reasonably controlled. 

The narrowest PSF appears on the optical axis and has a 90\% encircled power radius ($r_{\rm psf}$) less than 1\arcsec.  The PSF size increases and the effective area decreases with increasing off-axis angle. To utilize the high sensitivity of Chandra, we only consider CCD regions where $r_{\rm psf} < 5\arcsec$. The PSF radius as a function of off-axis angle is obtained from calibrations\footnote{We adopt the circular PSF radius for 90\% enclosed counts fraction at 1 keV, available at \url{https://asc.harvard.edu/cal/Hrma/PSFCore.html}.}.  This limits our search in a circular region with a radius of roughly 6\arcmin\ in each observation, on CCDs S2-S3 or I0-I3.

The frame time ($t_{\rm frm}$) of ACIS is 3.2~s nominally and adjustable depending on the number of CCDs in operation and the subarray mode. In the data, it ranges from 0.1 to 3.3~s, while the majority of observations (86\% of the total exposure) have a frame time of 3.1 or 3.2~s, i.e., with I0-I3 or S2-S3 in operation at the full frame mode. As the same search should be done in observations under the same condition (i.e., the same frame time, background level, and field of view), we thereby only consider observations in such configurations. In Table~\ref{tab:config}, different ACIS configurations along with the total number of frames in each configuration are listed. ACIS-S2 and S3 are treated separately because they have different background levels. Although we divide the search into different configurations, we note that all data should be considered as a whole for the control of false positives.

We adopt Chandra ACIS non-grating imaging observations in the timed exposure mode as of 2021 December 31 for the search. In sum, there are 11626 observations with a total exposure of 285~Ms in the configurations listed in Table~\ref{tab:config}. We conduct the search with level-2 events files in the energy range of 0.5--7~keV.  Level-1 events are used for removal of comic ray contaminations. The CIAO version 4.14 with CALDB 4.9.6 is used. 

Our first step is thus to identify clusters of photons ($n_{\rm ph} \ge 3$) in single frames.  We extract photons frame by frame based on the {\tt expno} (exposure number of CCD frame) column in the events file. The centroid of the photon positions is regarded as their central position, and the 90\% radius assuming a Rayleigh distribution is computed and defined as the cluster size, $r_{\rm clu} = [\ln(10) (\sigma_x^2 + \sigma_y^2)]^{1/2}$, where $\sigma_x$ and $\sigma_y$ are the standard deviation (root mean square around the average) along $x$ and $y$ direction, respectively.  We calculate the initial $r_{\rm clu}$ with all photons in the frame and then discard those outside $r_{\rm clu}$ and iterate until the remaining photons are all within $r_{\rm clu}$. With MARX simulations, we find that $r_{\rm clu}$ obtained with 3 photons has a mean well consistent with that constructed from many photons. Because the spatial distribution of source photons on the CCD plane follows the PSF, we consider a candidate transient event if $r_{\rm clu} \le r_{\rm psf}$.  

The choice of $n_{\rm ph} \ge 3$ is based on the assumption of Poisson distribution around the quiescent background level. If there is an X-ray source, the likelihood of seeing 3 or more photons within a PSF in a single frame will be enhanced. Also in that case, the way to evaluate the chance probability over the search region (6\arcmin\ around the optical axis) will be different and complicated, because the Poisson mean local to the event region is different from (much higher than) that in a pure background region. Therefore, we discard events if they appear on an X-ray source, based on the Chandra Source Catalog 2.0 \citep{Evans2010}, which covers observations up to 2019 October, and a local count comparison (to see if counts in $r_{\rm psf}$ are at least 5$\sigma$ in excess of the local background estimated from an annulus from $r_{\rm psf}$ to 15\arcsec, or a nearby source-free region). A different algorithm is needed to search for fast variations in an X-ray source \citep[e.g.,][]{Yang2019,Quirola-Vasquez2022}.  

\subsection{Discrimination of cosmic rays and background flares}
\label{sec:contamination}

Cosmic rays may penetrate the detector, resulting in ionization and energy deposits that appear like multiple photon events.  For each CCD event, a grade is assigned based on the pixel pattern, and events with certain grade numbers (0, 2, 3, 4, \& 6) are defined to be valid X-ray events. However, a small fraction of the cosmic ray events cannot be distinguished with the event grade.  Due to the high energies of these particles, they tend to produce straight ionization tracks.  Therefore, we check the level-1 data which contain events with all grade numbers to see if the transient candidate actually lies along a linear pattern together with invalid events (those assigned with a bad grade number) that are not shown in the level-2 data. If the level-1 events (both valid and invalid) outside the cluster but within 50\arcsec\ and the events in the cluster appear on the same linear pattern, we identify it as a cosmic ray event. 

Cosmic rays may also interact with surrounding materials and produce secondary particles with multiple hits in the CCD. In this case, the CCD events do not necessarily display a linear pattern, but may appear as a large cluster.  Thus, we assign a rank number to each transient candidate based on the number of level-1 events appearing beyond $r_{\rm clu}$ and within 50\arcsec, with rank = 0 for zero such event, rank = 1 for 1-2 events, and rank = 2 for 3-4 events. Those having 5 or more events around are discarded.  Transients with rank = 0 represent the most reliable identifications. In this work, as the first paper of this new method, we adopt the most conservative strategy and only present events with rank = 0, which naturally exclude cosmic ray events mentioned above. Events with rank > 0 will be discussed in the future. 

The background may exhibit flaring activity. During a background flare, the whole CCD will show an enhanced count rate.  We extract a lightcurve of the whole CCD at energies above 10 keV with a time step of 100~s, and identify time intervals for background flares as those with a count rate exceeding the quiescent level over $3\sigma$.  If a candidate transient event falls in the background flaring time interval, we discard it. 

The validation and verification report for each observation is examined. ObsID 581 may have issues with some CCD nodes and is discarded.  We also examined the locations of detected events to make sure that they are not due to hot pixels.

\subsection{Significance and control of false positives}
\label{sec:sig}
 
We calculate the significance for each candidate transient event in the detected frame.  The local mean count rate is estimated from an annulus with radii between $r_{\rm clu}$ and 15\arcsec, or a nearby source-free circular region if there are X-ray sources in the annulus.  Given a constant X-ray flux, the probability of seeing $n_{\rm ph}$ photons within $r_{\rm clu}$ is calculated assuming Poisson distribution. Then, it is divided by the fraction of the cluster size over the search region (field of view in Table~\ref{tab:config}), resulting in the chance probability of the event in that frame. A valid transient event is selected if the chance probability is less than a threshold that is small enough to rule out false positives. 

We adopt the Benjamini-Hochberg procedure \citep{Benjamini1995} to control the false discovery rate (FDR) for events from each configuration.  Given the total number of tests ($n_{\rm frm}$ in Table~\ref{tab:config}) and a FDR threshold $\alpha = 0.005$, there are $R = 11$ events in total with a probability $p < p_{\rm th}$ (listed in Table~\ref{tab:config}) that can be declared to be discoveries. By definition, the expected number of false positives is less than $\alpha R = 0.055$ summed over all configurations. This is also consistent with the simple estimate by multiplying the total number of frames with the threshold $p$-value in each configuration. The number of declared events $R$ depends on $\alpha$; here we have chosen an $\alpha$ such that the number of false positives in the final list is $<0.1$ and can be negligible.

Here we have assumed that the background level is constant over time and CCD locations, which is certainly not true. The total background (events with all grades and energies up to 15~keV) shows a time variation\footnote{\url{https://cxc.harvard.edu/cal/Acis/detailed\_info.html\#background}} by a factor of 2 \citep[also see][]{Suzuki2021}. The particle induced background, both the continuum and line components, displays a spatial variation with a much smaller factor\footnote{\url{https://cxc.harvard.edu/cal/Acis/Bkgrnd\_Prods/nonuniformity/acisbg.html}} \citep{Bartalucci2014,Suzuki2021}. If we assume that the good-grade background in the energy range of our interest has the same variation amplitude, it suggests that the $p$-value is under or overestimated by a factor of 3, which has a small impact on our results. In fact, our search algorithm disfavors candidates associated with a higher local background, because they have a higher $p$-value and a less chance to survive the FDR control. Thus, the presence of extended sources (such as clusters of galaxies) in some observations does not increase the false positives.

\begin{figure}
\centering
\includegraphics[width=\columnwidth]{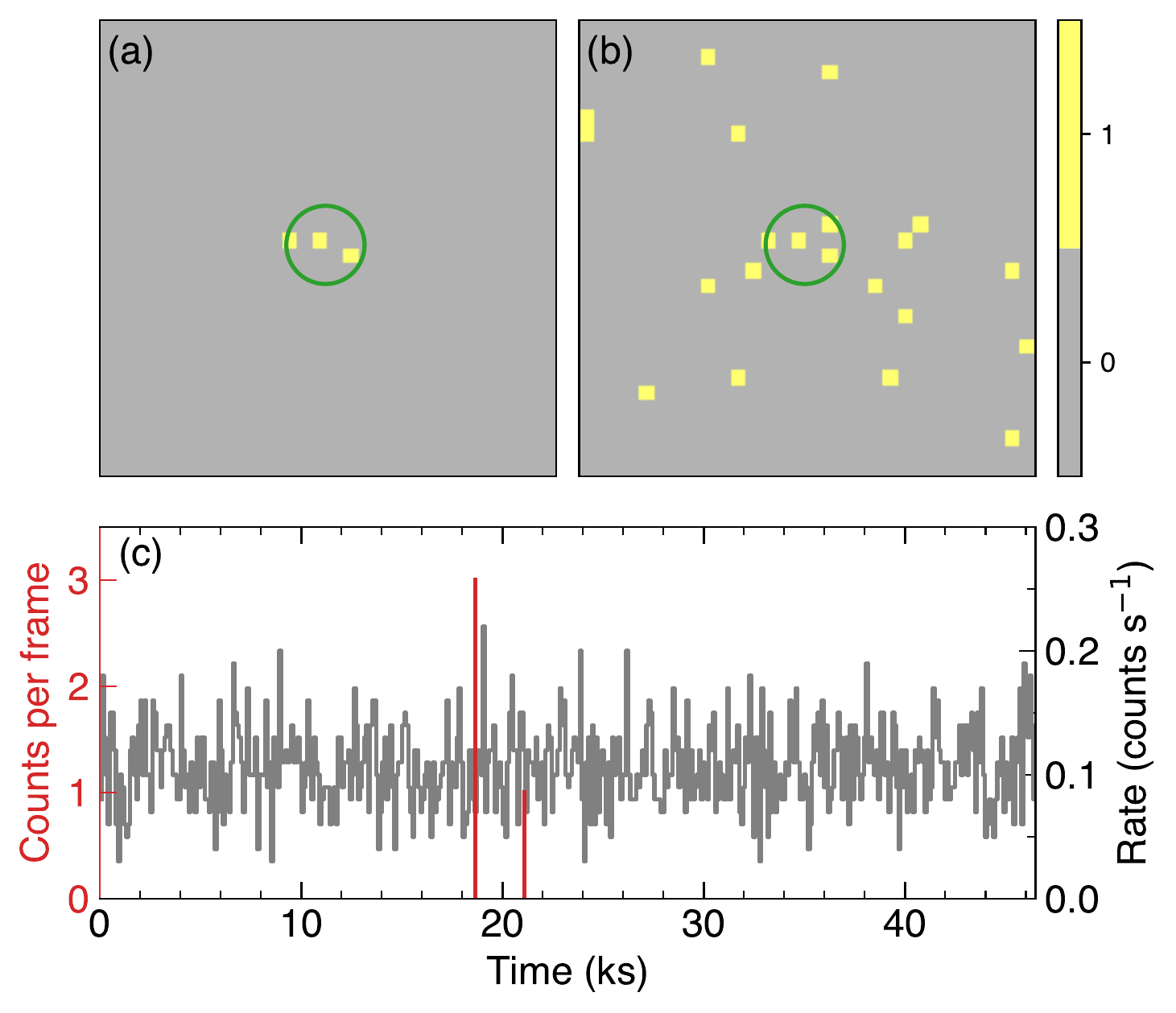}
\caption{Chandra X-ray images and lightcurves for RXT 130305 ({\bf a}) 0.5--7 keV image during the detection frame. The circle is centered on the event centroid with a radius of $r_{\rm clu}$.  ({\bf b}) 0.5--7 keV image integrated over the whole observation. ({\bf c}) Lightcurve for 0.5--7 keV photons inside the circle binned at the frame time ({\bf red}), and that for photons above 10~keV over the whole CCD binned at 100~s ({\bf gray}).  
\label{fig:search}}
\end{figure}

\section{Candidates and counterparts}
\label{sec:result}

\begin{figure}
\centering
\includegraphics[width=\columnwidth]{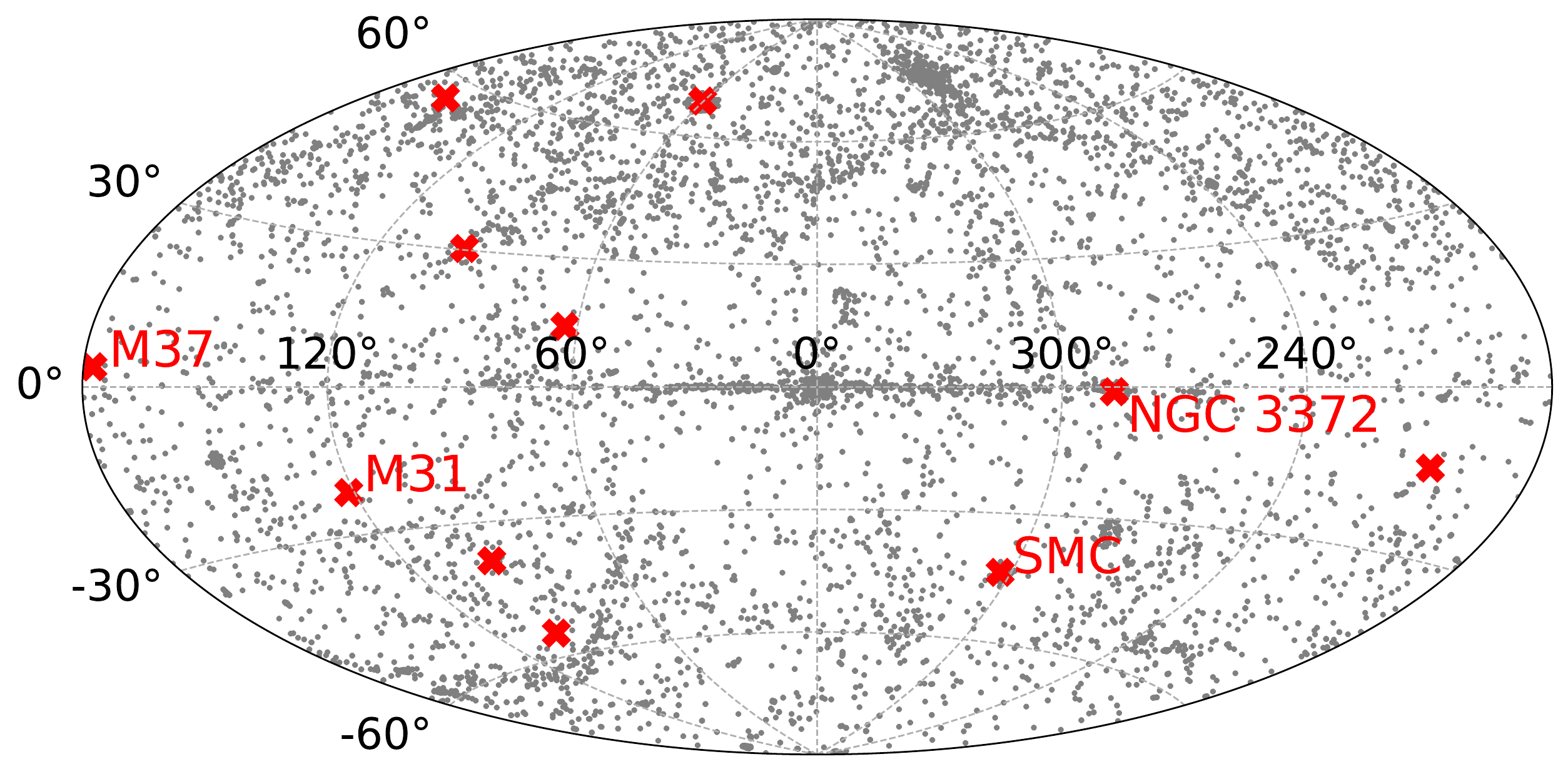}
\caption{Locations of the candidate rapid X-ray transients in Galactic coordinates. The gray points are Chandra observations used in the search. The possible host galaxies are labeled. 
\label{fig:loc}}
\end{figure}

After discrimination of cosmic rays and background flares, and control of FDR, we identified 11 candidate rapid transient events.  An example is shown in Figure~\ref{fig:search} to illustrate how the event looks like. We define the peak count rate of the transient events as $n_{\rm ph} / t_{\rm frm}$, and convert it to 0.3--8 keV flux assuming an absorbed power-law spectrum with $N_{\rm H} = 10^{21}$~cm$^{-2}$ and $\Gamma = 2$.  The conversion is based on local response files created using the {\tt specextract} task. We note that such a peak flux may have a high uncertainty (close to an order of magnitude) due to Poisson fluctuation with 3--4 counts. Thus, it is only useful for an order-of-magnitude estimate. The properties of the candidate events are listed in Table~\ref{tab:rxt}, including the time, sky location, detection significance, peak flux and rank, etc. Their sky locations in Galactic coordinates are plotted in Figure~\ref{fig:loc}. With MARX simulations, the position error determined with 3--4 photons is estimated to be less than 1\arcsec, 2\arcsec, and 3\arcsec, respectively, with $r_{\rm psf} = $ 1\arcsec, 3\arcsec, and 5\arcsec.

\begin{table*}
\footnotesize
\centering
\caption{Single-frame rapid X-ray transients detected with Chandra. }
\label{tab:rxt}
\begin{threeparttable}
\begin{tabular}{lccccccclccccc}
\hline\noalign{\smallskip}
No. & RXT & $t_0$ & R.A. & Decl. & $n_{\rm ph}$ & CCD & $t_{\rm frm}$ & $\log p$ & $r_{\rm clu}$ & $r_{\rm psf}$ & $\log f$ & ObsID \\ 
 &  & (UTC) & (J2000) & (J2000) &  &  & (s) &  & (\arcsec) & (\arcsec) &  & \\ 
(1) & (2) & (3) & (4) & (5) & (6) & (7) & (8) & (9) & (10) & (11) & (12) & (13)  \\ 
\noalign{\smallskip}\hline\noalign{\smallskip}
1 & 020823 & 2002-08-23T14:07:41.7 & 23:22:23.0 & $+$19:43:09 & 4 & S3 & 3.2 & $-$15.5 & 1.3 & 3.0 & $-$11.0 & 3028\\ 
2 & 060523 & 2006-05-23T06:29:50.2 & 23:57:46.4 & $+$05:13:20 & 3 & S3 & 3.2 & $-$9.3 & 1.4 & 2.0 & $-$11.1 & 7330\\ 
3 & 070602 & 2007-06-02T21:06:15.8 & 00:44:15.2 & $+$41:22:04 & 3 & I & 3.2 & $-$10.3 & 1.8 & 2.9 & $-$10.4 & 7067\\ 
4 & 080723 & 2008-07-23T00:37:28.9 & 10:40:35.3 & $-$59:55:52 & 3 & I & 3.2 & $-$8.8 & 2.2 & 4.8 & $-$10.4 & 9497\\ 
5 & 081128 & 2008-11-28T01:23:13.7 & 05:46:24.2 & $-$00:07:57 & 3 & I & 3.2 & $-$10.3 & 1.6 & 3.8 & $-$10.3 & 10763\\ 
6 & 090128 & 2009-01-28T05:52:40.1 & 10:49:31.0 & $+$51:07:42 & 4 & I & 3.2 & $-$13.4 & 3.0 & 3.3 & $-$10.6 & 10534\\ 
7 & 110123 & 2011-01-23T02:49:54.3 & 18:53:07.2 & $+$33:02:01 & 3 & S3 & 3.2 & $-$9.2 & 1.4 & 4.8 & $-$11.0 & 12364\\ 
8 & 110316 & 2011-03-16T13:00:37.8 & 17:55:00.6 & $+$66:35:17 & 3 & I & 3.2 & $-$9.6 & 1.7 & 2.3 & $-$10.7 & 12931\\ 
9 & 111115 & 2011-11-15T06:14:30.2 & 05:52:43.8 & $+$32:32:58 & 3 & I & 3.1 & $-$11.3 & 1.0 & 4.3 & $-$9.9 & 13655\\ 
10 & 130305 & 2013-03-05T04:05:32.3 & 00:47:26.3 & $-$73:13:13 & 3 & I & 3.1 & $-$10.1 & 1.3 & 4.2 & $-$10.1 & 14674\\ 
11 & 140323 & 2014-03-23T21:22:13.7 & 14:26:01.7 & $+$35:08:23 & 3 & I & 3.1 & $-$10.3 & 1.6 & 3.6 & $-$10.6 & 16321\\ 
\noalign{\smallskip}\hline\noalign{\smallskip}
\end{tabular}
\begin{tablenotes}
\footnotesize
\item {\bf Note.}
Column~1: Object number.
Column~2: Event name following the GRB naming convention. 
Column~3: Right ascension. 
Column~4: Declination. 
Column~5: Arrival time of the event.
Column~6: Number of photons.
Column~7: CCD chip (ACIS-I, ACIS-S2, or ACIS-S3).
Column~8: CCD frame time.
Column~9: Logarithmic chance probability of the transient event.
Column~10: Cluster size.
Column~11: Radius of the local PSF that encircles 90\% counts.
Column~12: Log of the observed peak flux in units of \ergcms, converted from the peak count rate $n_{\rm ph}/t_{\rm frm}$ assuming a power-law spectrum with $N_{\rm H} = 10^{21}$~cm$^{-2}$ and $\Gamma = 2$.
Column~13: Chandra observation ID. 
\end{tablenotes}
\end{threeparttable}
\end{table*}

\subsection{Counterparts}
\label{sec:cp}

The transient location is calculated as the centroid of the few (3--4) photons in the detection frame.  The 90\% absolute astrometric accuracy\footnote{\url{https://cxc.harvard.edu/cal/ASPECT/celmon/}} for Chandra sources is 0\farcs8.  All of our events are found at relatively large off-axis angles where $r_{\rm psf}$ is larger than 0\farcs8. With MARX simulations, we find that the position uncertainty with 3 photons is smaller than $r_{\rm psf}$. Thus, to be conservative, we adopt $r_{\rm psf}$ as the error radius.

Given the Chandra coordinates and error radius mentioned above, we search for their optical counterparts or hosts in published astronomical catalogs. We adopt the GRB naming convention in designation of sources. Possible associations with host star clusters or galaxies are found for 4 transients.

\begin{itemize}
\itemsep0em

\item \textit{RXT 070602}. It appears along the line of sight to the disk of M31, which has a distance of 0.78~Mpc \citep{Bhardwaj2016}.

\item \textit{RXT 080723}. It is located along the direction to the Carina Nebula (NGC 3372), which has a distance of 2.6~kpc \citep{Kuhn2019}.

\item \textit{RXT 111115}. It appears on the outskirt of the open cluster M37, which has a distance of 1.4~kpc \citep{Cantat-Gaudin2020}. 

\item \textit{RXT 130305}. It is spatially associated with the Small Magellanic Cloud (SMC), which has a distance of 62~kpc \citep{Graczyk2020}.

\end{itemize}

Possible optical counterparts are found for 2 transients. The chance probability of coincidence is calculated based on the source density within 2\arcmin\ around the event and the size of the match circle (with a radius of $r_{\rm psf}$).

\begin{itemize}
\itemsep0em

\item \textit{RXT 110123}. An optical counterpart is found in Pan-STARRS DR1 at separation of 4\farcs5, with $r=21.8$, $i=21.1$, and $z=20.6$. The probability of chance coincidence is 0.29.

\item \textit{RXT 140323}. An object at a separation of 2\farcs3 with a photometric redshift $z_{\rm phot} = 2.256$ is found in the Bo\"{o}tes Field Photometric redshift catalog \citep{Duncan2021}. The probability of chance coincidence is 0.45.

\end{itemize}

We note that, based on the probability of chance coincidence, both associations should be treated with cautions. Besides the above possible associations, there are another 4 objects (RXTs 020823, 060523, 090128, \& 110316) located in the Pan-STARRS DR1 field, without any detectable optical counterparts within their error radius. This allows us to set upper limits on their optical flux as $\sim$22~mag in $gri$ and 20--21~mag in $zy$, by finding the faintest Pan-STARRS object within a 2\arcmin\ vicinity.

\section{Discussion}
\label{sec:discuss}

We identified 11 single-frame rapid X-ray transients in the Chandra archive from regions without point-like sources, and the expected number of false positives is controlled to be negligible ($< 0.1$). Among them, two are spatially coincident with a Galactic extended object (RXTs 080723 \& 111115); one may have a SMC origin (RXT 130305); one has a possible association with M31 (RXT 070602); the rest have no or a dim optical counterpart. Their sky locations seem random, or at least, not clustered on the Galactic plane (see Figure~\ref{fig:loc}), suggesting that many of them are not originated in the Milky Way.

Assuming that the RXTs with no bright optical counterparts or nearby associations have a cosmological origin, we may translate their measured peak fluxes, which are around $10^{-11}$~\ergcms, to peak luminosities. Given a standard cosmology ($h = 0.7$, $\Omega_m = 0.3$, and $\Omega_\Lambda = 0.7$), the rest-frame 0.3--8 keV band luminosity is estimated to be $1 \times 10^{45}$, $5 \times 10^{46}$, and $3 \times 10^{47}$~\ergs, at a redshift of 0.2, 1, and 2, respectively. The uncertainty can be estimated based on a Bayesian approach assuming Poisson distribution given the source and background counts using the {\tt aprates} task in CIAO, and is found to be about 0.4~dex at 90\% credible level.

Such a high peak luminosity indicates that they may be energetic bursts like GRBs or due to neutron star merges.  The isotropic peak luminosity of the prompt GRB emission ranges from $10^{47}$ to $10^{54}$~\ergs\ in the 10--10000 keV band \citep{Zhang2009}. Assuming a typical GRB spectrum, i.e., a band spectrum ({\tt grbm} in XSPEC) with $\alpha = -1$, $\beta = -2.3$, and $E_{\rm c} = 200$~keV \citep{Preece2000}, the 0.3--8 keV luminosity is found to be roughly in the range of $10^{45} - 10^{52}$~\ergs. A long-duration GRB with a sharp spike or a short-duration GRB can thus produce a rapid transient event in a CCD frame of Chandra.  As almost every long GRB is supposed to be followed by an X-ray afterglow that can be detected with Chandra \citep{Gehrels2009}, short GRBs seem more likely to be responsible for the transient events. We argue that it is unlikely that these transient events are orphan afterglows with the prompt GRB emission missed due to large off-axis viewing \citep{Sarin2021}, as the emission may last much longer than a few seconds before decaying below the detection sensitivity of Chandra. 

The expected number of RXTs due to short GRBs can be estimated as
\begin{equation}
N = \rho \frac{{\rm FoV}}{4 \pi} t_{\rm exp} \frac{4 \pi D^3}{3} \, ,
\end{equation}
where $\rho$ is the event rate density, FoV is the searching solid angle roughly of 100 arcmin$^2$ (Table~\ref{tab:config}; note that S2 and S3 are usually in operation simultaneously but are listed separately), $t_{\rm exp}$ is the total exposure time (285~Ms), and $D$ is the horizon distance. For short GRBs above $10^{50}$~\ergs\ in 1--$10^4$~keV, or $\sim 10^{48}$~\ergs\ in 0.3--8 keV, the horizon distance $D \approx 30$~Gpc and event rate density $\rho = 3.3_{-0.8}^{+1.0}$~Gpc$^{-3}$~yr$^{-1}$ \citep{Sun2015}. Thus, the expected detection number $N = 2.3_{-0.6}^{+0.7}$, suggesting that some of our RXTs may be due to such events. For 170817A-like GRBs, which has a high event rate density but a low luminosity \citep{Mandhai2018,Zhang2018} and correspondingly short horizon (1.7~Gpc), we do not expect to detect similar events ($N = 0.024_{-0.020}^{+0.055}$) in the current Chandra archive.

Merger of binary neutron stars may produce a differential rotation supported neutron star with transient electromagnetic emission \citep{Zhang2013,Sun2017,Sun2019}. These events may last $10^2 - 10^3$~s \citep[e.g.,][]{Bauer2017,Xue2019}. However, if it produces a hypermassive neutron star \citep{Shapiro2000}, which can survive tens to hundreds of milliseconds before collapsing into a black hole, the flare may be detected as a single-frame rapid transient.

For RXTs 070602 (M31), 080723 (NGC 3372), 111115 (M37), \& 130305 (SMC), if we assume that they are indeed associated with their hosts (shown in brackets), the inferred peak luminosity is $3_{-2}^{+4} \times 10^{39}$, $3_{-2}^{+4} \times 10^{34}$, $3_{-2}^{+4} \times 10^{34}$, $4_{-3}^{+5} \times 10^{37}$~\ergs, respectively in the same order. These are consistent with the luminosities that can be produced by Galactic X-ray binaries \citep{Remillard2006} and ultraluminous X-ray sources \citep{Kaaret2017}.

The lack of detection with Chandra for persistent emission and faintness of their optical counterparts may also suggest that they could be Galactic low-mass X-ray binaries in the quiescent state. X-ray binaries and in particular short-period black hole binaries in the quiescent state can have a persistent luminosity as low as $3 \times 10^{30}$~\ergs\ \citep{Menou1999}. When the accretion rate is rather low, it is difficult to identify them even with the most sensitive X-ray telescope. If they exhibit short duration flares in the quiescent state, they may be detected as rapid X-ray transients on top of pure background. Recently, a lot of efforts have been made to search for quiescent black holes using optical dynamics \citep[e.g.,][]{Giesers2018,Thompson2019,Liu2019}. Therefore, one may check their single-frame variability in the X-ray data with focusing telescopes using our method before concluding that there is no detectable X-ray emission from the source position.    If we assume a minimum variability timescale of $10 G M_{\rm BH} / c^3$, the method allows us to detect black holes with $M_{\rm BH} < 6.3 \times 10^4$~$M_{\sun}$ given a frame time of 3.1--3.2~s.  

Magnetars are another possibility \citep{Kaspi2017}. The persistent X-ray luminosity for some magnetars could be well below $10^{33}$~\ergs\ and even Chandra cannot detect them \citep[e.g.,][]{Mori2013,Zhou2014}.  They could produce short flares with a sub-second duration and a total energy in the range of $10^{36}-10^{39}$~erg \citep[e.g.,][]{Goegues2001}.  Those flares, if caught by Chandra, will appear like our rapid transients. Giant flares from magnetars may exhibit a major, sub-second spike with a peak luminosity up to $10^{47}$~\ergs, followed by pulsed, relatively fainter emission \citep[e.g.,][]{Hurley2005}. If such a giant flare is located at a cosmological distance, it can be detected as a single-frame rapid transient with Chandra.  

Late-type stars may produce stellar flares due to magnetic activities \citep[e.g.,][]{Nordon2007,Kuznetsov2021}. They could be optically dim and potentially contribute to our detections. However, the flare timescale typically ranges from minutes to hours \citep{Benz2017}, much longer than what we have observed. Thus, this scenario is possible only if the flares are produced due to extreme magnetic fields or arise from very compact regions \citep{Maehara2015,Namekata2017}.

To conclude, we proposed an algorithm that allows us to identify single-frame rapid X-ray transients detected with focusing X-ray telescopes. They are second-duration faint flares, without persistent X-ray emission.  Based on ACIS imaging observations in the public Chandra archive, we found 11 candidates and argued that they are unlikely due to cosmic rays or background flares. The nature of these transients is unknown. They could be short GRBs, hypermassive short-lived neutron stars, X-ray binaries in the quiescent state, magnetars, or stellar flares.  Future in-depth studies in both the X-ray and multiwavelength bands are important for us to understand these events. 

\section*{Acknowledgements}

We thank Catherine Grant, Bing Zhang, Bin-Bin Zhang, and Hui Tian for helpful discussions, and the anonymous referee for useful comments. HF acknowledges funding support from the National Key R\&D Project under grant 2018YFA0404502, the National Natural Science Foundation of China under grants Nos.\ 12025301, 12103027, \& 11821303, and the Tsinghua University Initiative Scientific Research Program.


\begin{thebibliography}{}
\makeatletter
\relax
\def\mn@urlcharsother{\let\do\@makeother \do\$\do\&\do\#\do\^\do\_\do\%\do\~}
\def\mn@doi{\begingroup\mn@urlcharsother \@ifnextchar [ {\mn@doi@}
  {\mn@doi@[]}}
\def\mn@doi@[#1]#2{\def\@tempa{#1}\ifx\@tempa\@empty \href
  {http://dx.doi.org/#2} {doi:#2}\else \href {http://dx.doi.org/#2} {#1}\fi
  \endgroup}
\def\mn@eprint#1#2{\mn@eprint@#1:#2::\@nil}
\def\mn@eprint@arXiv#1{\href {http://arxiv.org/abs/#1} {{\tt arXiv:#1}}}
\def\mn@eprint@dblp#1{\href {http://dblp.uni-trier.de/rec/bibtex/#1.xml}
  {dblp:#1}}
\def\mn@eprint@#1:#2:#3:#4\@nil{\def\@tempa {#1}\def\@tempb {#2}\def\@tempc
  {#3}\ifx \@tempc \@empty \let \@tempc \@tempb \let \@tempb \@tempa \fi \ifx
  \@tempb \@empty \def\@tempb {arXiv}\fi \@ifundefined
  {mn@eprint@\@tempb}{\@tempb:\@tempc}{\expandafter \expandafter \csname
  mn@eprint@\@tempb\endcsname \expandafter{\@tempc}}}

\bibitem[\protect\citeauthoryear{{Abbott} et~al.,}{{Abbott}
  et~al.}{2017}]{Abbott2017}
{Abbott} B.~P.,  et~al., 2017, \mn@doi [\apjl] {10.3847/2041-8213/aa920c},
  \href {https://ui.adsabs.harvard.edu/abs/2017ApJ...848L..13A} {848, L13}

\bibitem[\protect\citeauthoryear{{Bartalucci}, {Mazzotta}, {Bourdin}  \&
  {Vikhlinin}}{{Bartalucci} et~al.}{2014}]{Bartalucci2014}
{Bartalucci} I.,  {Mazzotta} P.,  {Bourdin} H.,   {Vikhlinin} A.,  2014,
  \mn@doi [\aap] {10.1051/0004-6361/201423443}, \href
  {https://ui.adsabs.harvard.edu/abs/2014A&A...566A..25B} {566, A25}

\bibitem[\protect\citeauthoryear{{Bauer} et~al.,}{{Bauer}
  et~al.}{2017}]{Bauer2017}
{Bauer} F.~E.,  et~al., 2017, \mn@doi [\mnras] {10.1093/mnras/stx417}, \href
  {https://ui.adsabs.harvard.edu/abs/2017MNRAS.467.4841B} {467, 4841}

\bibitem[\protect\citeauthoryear{Benjamini \& Hochberg}{Benjamini \&
  Hochberg}{1995}]{Benjamini1995}
Benjamini Y.,  Hochberg Y.,  1995, Journal of the Royal Statistical Society.
  Series B (Methodological), 57, 289

\bibitem[\protect\citeauthoryear{{Benz}}{{Benz}}{2017}]{Benz2017}
{Benz} A.~O.,  2017, \mn@doi [Living Reviews in Solar Physics]
  {10.1007/s41116-016-0004-3}, \href
  {https://ui.adsabs.harvard.edu/abs/2017LRSP...14....2B} {14, 2}

\bibitem[\protect\citeauthoryear{{Bhardwaj}, {Kanbur}, {Macri}, {Singh},
  {Ngeow}, {Wagner-Kaiser}  \& {Sarajedini}}{{Bhardwaj}
  et~al.}{2016}]{Bhardwaj2016}
{Bhardwaj} A.,  {Kanbur} S.~M.,  {Macri} L.~M.,  {Singh} H.~P.,  {Ngeow} C.-C.,
   {Wagner-Kaiser} R.,   {Sarajedini} A.,  2016, \mn@doi [\aj]
  {10.3847/0004-6256/151/4/88}, \href
  {https://ui.adsabs.harvard.edu/abs/2016AJ....151...88B} {151, 88}

\bibitem[\protect\citeauthoryear{{Boella}, {Butler}, {Perola}, {Piro}, {Scarsi}
   \& {Bleeker}}{{Boella} et~al.}{1997}]{Boella1997}
{Boella} G.,  {Butler} R.~C.,  {Perola} G.~C.,  {Piro} L.,  {Scarsi} L.,
  {Bleeker} J.~A.~M.,  1997, \mn@doi [\aaps] {10.1051/aas:1997136}, \href
  {https://ui.adsabs.harvard.edu/abs/1997A&AS..122..299B} {122, 299}

\bibitem[\protect\citeauthoryear{{Cantat-Gaudin} \& {Anders}}{{Cantat-Gaudin}
  \& {Anders}}{2020}]{Cantat-Gaudin2020}
{Cantat-Gaudin} T.,  {Anders} F.,  2020, \mn@doi [\aap]
  {10.1051/0004-6361/201936691}, \href
  {https://ui.adsabs.harvard.edu/abs/2020A&A...633A..99C} {633, A99}

\bibitem[\protect\citeauthoryear{{Cenko} et~al.,}{{Cenko}
  et~al.}{2010}]{Cenko2010}
{Cenko} S.~B.,  et~al., 2010, \mn@doi [\apj] {10.1088/0004-637X/711/2/641},
  \href {https://ui.adsabs.harvard.edu/abs/2010ApJ...711..641C} {711, 641}

\bibitem[\protect\citeauthoryear{{Chakrabarty}, {Morgan}, {Muno}, {Galloway},
  {Wijnands}, {van der Klis}  \& {Markwardt}}{{Chakrabarty}
  et~al.}{2003}]{Chakrabarty2003}
{Chakrabarty} D.,  {Morgan} E.~H.,  {Muno} M.~P.,  {Galloway} D.~K.,
  {Wijnands} R.,  {van der Klis} M.,   {Markwardt} C.~B.,  2003, \mn@doi [\nat]
  {10.1038/nature01732}, \href
  {https://ui.adsabs.harvard.edu/abs/2003Natur.424...42C} {424, 42}

\bibitem[\protect\citeauthoryear{{Duncan} et~al.,}{{Duncan}
  et~al.}{2021}]{Duncan2021}
{Duncan} K.~J.,  et~al., 2021, \mn@doi [\aap] {10.1051/0004-6361/202038809},
  \href {https://ui.adsabs.harvard.edu/abs/2021A&A...648A...4D} {648, A4}

\bibitem[\protect\citeauthoryear{{Evans} et~al.,}{{Evans}
  et~al.}{2010}]{Evans2010}
{Evans} I.~N.,  et~al., 2010, \mn@doi [\apjs] {10.1088/0067-0049/189/1/37},
  \href {https://ui.adsabs.harvard.edu/abs/2010ApJS..189...37E} {189, 37}

\bibitem[\protect\citeauthoryear{{Fishman} et~al.,}{{Fishman}
  et~al.}{1993}]{Fishman1993}
{Fishman} G.~J.,  et~al., 1993, \aaps, \href
  {https://ui.adsabs.harvard.edu/abs/1993A&AS...97...17F} {97, 17}

\bibitem[\protect\citeauthoryear{{Fishman} et~al.,}{{Fishman}
  et~al.}{1994}]{Fishman1994}
{Fishman} G.~J.,  et~al., 1994, \mn@doi [Science]
  {10.1126/science.264.5163.1313}, \href
  {https://ui.adsabs.harvard.edu/abs/1994Sci...264.1313F} {264, 1313}

\bibitem[\protect\citeauthoryear{{Gehrels} et~al.,}{{Gehrels}
  et~al.}{2004}]{Gehrels2004}
{Gehrels} N.,  et~al., 2004, \mn@doi [\apj] {10.1086/422091}, \href
  {https://ui.adsabs.harvard.edu/abs/2004ApJ...611.1005G} {611, 1005}

\bibitem[\protect\citeauthoryear{{Gehrels}, {Ramirez-Ruiz}  \& {Fox}}{{Gehrels}
  et~al.}{2009}]{Gehrels2009}
{Gehrels} N.,  {Ramirez-Ruiz} E.,   {Fox} D.~B.,  2009, \mn@doi [\araa]
  {10.1146/annurev.astro.46.060407.145147}, \href
  {https://ui.adsabs.harvard.edu/abs/2009ARA&A..47..567G} {47, 567}

\bibitem[\protect\citeauthoryear{{Giesers} et~al.,}{{Giesers}
  et~al.}{2018}]{Giesers2018}
{Giesers} B.,  et~al., 2018, \mn@doi [\mnras] {10.1093/mnrasl/slx203}, \href
  {https://ui.adsabs.harvard.edu/abs/2018MNRAS.475L..15G} {475, L15}

\bibitem[\protect\citeauthoryear{{G{\"o}{\v{g}}{\"u}{\c{s}}}, {Kouveliotou},
  {Woods}, {Thompson}, {Duncan}  \& {Briggs}}{{G{\"o}{\v{g}}{\"u}{\c{s}}}
  et~al.}{2001}]{Goegues2001}
{G{\"o}{\v{g}}{\"u}{\c{s}}} E.,  {Kouveliotou} C.,  {Woods} P.~M.,  {Thompson}
  C.,  {Duncan} R.~C.,   {Briggs} M.~S.,  2001, \mn@doi [\apj]
  {10.1086/322463}, \href
  {https://ui.adsabs.harvard.edu/abs/2001ApJ...558..228G} {558, 228}

\bibitem[\protect\citeauthoryear{{Graczyk} et~al.,}{{Graczyk}
  et~al.}{2020}]{Graczyk2020}
{Graczyk} D.,  et~al., 2020, \mn@doi [\apj] {10.3847/1538-4357/abbb2b}, \href
  {https://ui.adsabs.harvard.edu/abs/2020ApJ...904...13G} {904, 13}

\bibitem[\protect\citeauthoryear{{Hurley} et~al.,}{{Hurley}
  et~al.}{2005}]{Hurley2005}
{Hurley} K.,  et~al., 2005, \mn@doi [\nat] {10.1038/nature03519}, \href
  {https://ui.adsabs.harvard.edu/abs/2005Natur.434.1098H} {434, 1098}

\bibitem[\protect\citeauthoryear{{Irwin} et~al.,}{{Irwin}
  et~al.}{2016}]{Irwin2016}
{Irwin} J.~A.,  et~al., 2016, \mn@doi [\nat] {10.1038/nature19822}, \href
  {https://ui.adsabs.harvard.edu/abs/2016Natur.538..356I} {538, 356}

\bibitem[\protect\citeauthoryear{{Kaaret}, {Feng}  \& {Roberts}}{{Kaaret}
  et~al.}{2017}]{Kaaret2017}
{Kaaret} P.,  {Feng} H.,   {Roberts} T.~P.,  2017, \mn@doi [\araa]
  {10.1146/annurev-astro-091916-055259}, \href
  {https://ui.adsabs.harvard.edu/abs/2017ARA&A..55..303K} {55, 303}

\bibitem[\protect\citeauthoryear{{Kaspi} \& {Beloborodov}}{{Kaspi} \&
  {Beloborodov}}{2017}]{Kaspi2017}
{Kaspi} V.~M.,  {Beloborodov} A.~M.,  2017, \mn@doi [\araa]
  {10.1146/annurev-astro-081915-023329}, \href
  {https://ui.adsabs.harvard.edu/abs/2017ARA&A..55..261K} {55, 261}

\bibitem[\protect\citeauthoryear{{Kawai} et~al.,}{{Kawai}
  et~al.}{1999}]{Kawai1999}
{Kawai} N.,  et~al., 1999, \mn@doi [\aaps] {10.1051/aas:1999353}, \href
  {https://ui.adsabs.harvard.edu/abs/1999A&AS..138..563K} {138, 563}

\bibitem[\protect\citeauthoryear{{Kuhn}, {Hillenbrand}, {Sills}, {Feigelson}
  \& {Getman}}{{Kuhn} et~al.}{2019}]{Kuhn2019}
{Kuhn} M.~A.,  {Hillenbrand} L.~A.,  {Sills} A.,  {Feigelson} E.~D.,   {Getman}
  K.~V.,  2019, \mn@doi [\apj] {10.3847/1538-4357/aaef8c}, \href
  {https://ui.adsabs.harvard.edu/abs/2019ApJ...870...32K} {870, 32}

\bibitem[\protect\citeauthoryear{{Kumar} \& {Zhang}}{{Kumar} \&
  {Zhang}}{2015}]{Kumar2015}
{Kumar} P.,  {Zhang} B.,  2015, \mn@doi [\physrep]
  {10.1016/j.physrep.2014.09.008}, \href
  {https://ui.adsabs.harvard.edu/abs/2015PhR...561....1K} {561, 1}

\bibitem[\protect\citeauthoryear{{Kuznetsov} \& {Kolotkov}}{{Kuznetsov} \&
  {Kolotkov}}{2021}]{Kuznetsov2021}
{Kuznetsov} A.~A.,  {Kolotkov} D.~Y.,  2021, \mn@doi [\apj]
  {10.3847/1538-4357/abf569}, \href
  {https://ui.adsabs.harvard.edu/abs/2021ApJ...912...81K} {912, 81}

\bibitem[\protect\citeauthoryear{{Li} et~al.,}{{Li} et~al.}{2020}]{Li2020}
{Li} Y.,  et~al., 2020, \mn@doi [Scientia Sinica Physica, Mechanica \&
  Astronomica] {10.1360/SSPMA-2019-0417}, \href
  {https://ui.adsabs.harvard.edu/abs/2020SSPMA..50l9508L} {50, 129508}

\bibitem[\protect\citeauthoryear{{Liu} et~al.,}{{Liu} et~al.}{2019}]{Liu2019}
{Liu} J.,  et~al., 2019, \mn@doi [\nat] {10.1038/s41586-019-1766-2}, \href
  {https://ui.adsabs.harvard.edu/abs/2019Natur.575..618L} {575, 618}

\bibitem[\protect\citeauthoryear{{Maehara}, {Shibayama}, {Notsu}, {Notsu},
  {Honda}, {Nogami}  \& {Shibata}}{{Maehara} et~al.}{2015}]{Maehara2015}
{Maehara} H.,  {Shibayama} T.,  {Notsu} Y.,  {Notsu} S.,  {Honda} S.,  {Nogami}
  D.,   {Shibata} K.,  2015, \mn@doi [Earth, Planets and Space]
  {10.1186/s40623-015-0217-z}, \href
  {https://ui.adsabs.harvard.edu/abs/2015EP&S...67...59M} {67, 59}

\bibitem[\protect\citeauthoryear{{Mandhai}, {Tanvir}, {Lamb}, {Levan}  \&
  {Tsang}}{{Mandhai} et~al.}{2018}]{Mandhai2018}
{Mandhai} S.,  {Tanvir} N.,  {Lamb} G.,  {Levan} A.,   {Tsang} D.,  2018,
  \mn@doi [Galaxies] {10.3390/galaxies6040130}, \href
  {https://ui.adsabs.harvard.edu/abs/2018Galax...6..130M} {6, 130}

\bibitem[\protect\citeauthoryear{{Meegan} et~al.,}{{Meegan}
  et~al.}{2009}]{Meegan2009}
{Meegan} C.,  et~al., 2009, \mn@doi [\apj] {10.1088/0004-637X/702/1/791}, \href
  {https://ui.adsabs.harvard.edu/abs/2009ApJ...702..791M} {702, 791}

\bibitem[\protect\citeauthoryear{{Menou}, {Esin}, {Narayan}, {Garcia}, {Lasota}
   \& {McClintock}}{{Menou} et~al.}{1999}]{Menou1999}
{Menou} K.,  {Esin} A.~A.,  {Narayan} R.,  {Garcia} M.~R.,  {Lasota} J.-P.,
  {McClintock} J.~E.,  1999, \mn@doi [\apj] {10.1086/307443}, \href
  {https://ui.adsabs.harvard.edu/abs/1999ApJ...520..276M} {520, 276}

\bibitem[\protect\citeauthoryear{{Mereghetti}}{{Mereghetti}}{2008}]{Mereghetti2008}
{Mereghetti} S.,  2008, \mn@doi [\aapr] {10.1007/s00159-008-0011-z}, \href
  {https://ui.adsabs.harvard.edu/abs/2008A&ARv..15..225M} {15, 225}

\bibitem[\protect\citeauthoryear{{Mori} et~al.,}{{Mori}
  et~al.}{2013}]{Mori2013}
{Mori} K.,  et~al., 2013, \mn@doi [\apjl] {10.1088/2041-8205/770/2/L23}, \href
  {https://ui.adsabs.harvard.edu/abs/2013ApJ...770L..23M} {770, L23}

\bibitem[\protect\citeauthoryear{{Namekata} et~al.,}{{Namekata}
  et~al.}{2017}]{Namekata2017}
{Namekata} K.,  et~al., 2017, \mn@doi [\apj] {10.3847/1538-4357/aa9b34}, \href
  {https://ui.adsabs.harvard.edu/abs/2017ApJ...851...91N} {851, 91}

\bibitem[\protect\citeauthoryear{{Nordon} \& {Behar}}{{Nordon} \&
  {Behar}}{2007}]{Nordon2007}
{Nordon} R.,  {Behar} E.,  2007, \mn@doi [\aap] {10.1051/0004-6361:20066449},
  \href {https://ui.adsabs.harvard.edu/abs/2007A&A...464..309N} {464, 309}

\bibitem[\protect\citeauthoryear{{Preece}, {Briggs}, {Mallozzi}, {Pendleton},
  {Paciesas}  \& {Band}}{{Preece} et~al.}{2000}]{Preece2000}
{Preece} R.~D.,  {Briggs} M.~S.,  {Mallozzi} R.~S.,  {Pendleton} G.~N.,
  {Paciesas} W.~S.,   {Band} D.~L.,  2000, \mn@doi [\apjs] {10.1086/313289},
  \href {https://ui.adsabs.harvard.edu/abs/2000ApJS..126...19P} {126, 19}

\bibitem[\protect\citeauthoryear{{Quirola-V{\'a}squez}
  et~al.,}{{Quirola-V{\'a}squez} et~al.}{2022}]{Quirola-Vasquez2022}
{Quirola-V{\'a}squez} J.,  et~al., 2022, \mn@doi [\aap]
  {10.1051/0004-6361/202243047}, \href
  {https://ui.adsabs.harvard.edu/abs/2022A&A...663A.168Q} {663, A168}

\bibitem[\protect\citeauthoryear{{Remillard} \& {McClintock}}{{Remillard} \&
  {McClintock}}{2006}]{Remillard2006}
{Remillard} R.~A.,  {McClintock} J.~E.,  2006, \mn@doi [\araa]
  {10.1146/annurev.astro.44.051905.092532}, \href
  {https://ui.adsabs.harvard.edu/abs/2006ARA&A..44...49R} {44, 49}

\bibitem[\protect\citeauthoryear{{Sarin}, {Ashton}, {Lasky}, {Ackley}, {Mong}
  \& {Galloway}}{{Sarin} et~al.}{2021}]{Sarin2021}
{Sarin} N.,  {Ashton} G.,  {Lasky} P.~D.,  {Ackley} K.,  {Mong} Y.-L.,
  {Galloway} D.~K.,  2021, \mn@doi [arXiv e-prints]
  {10.48550/arXiv.2105.10108}, \href
  {https://ui.adsabs.harvard.edu/abs/2021arXiv210510108S} {p. arXiv:2105.10108}

\bibitem[\protect\citeauthoryear{{Shapiro}}{{Shapiro}}{2000}]{Shapiro2000}
{Shapiro} S.~L.,  2000, \mn@doi [\apj] {10.1086/317209}, \href
  {https://ui.adsabs.harvard.edu/abs/2000ApJ...544..397S} {544, 397}

\bibitem[\protect\citeauthoryear{{Stanek} et~al.,}{{Stanek}
  et~al.}{2003}]{Stanek2003}
{Stanek} K.~Z.,  et~al., 2003, \mn@doi [\apjl] {10.1086/376976}, \href
  {https://ui.adsabs.harvard.edu/abs/2003ApJ...591L..17S} {591, L17}

\bibitem[\protect\citeauthoryear{{Sun}, {Zhang}  \& {Li}}{{Sun}
  et~al.}{2015}]{Sun2015}
{Sun} H.,  {Zhang} B.,   {Li} Z.,  2015, \mn@doi [\apj]
  {10.1088/0004-637X/812/1/33}, \href
  {https://ui.adsabs.harvard.edu/abs/2015ApJ...812...33S} {812, 33}

\bibitem[\protect\citeauthoryear{{Sun}, {Zhang}  \& {Gao}}{{Sun}
  et~al.}{2017}]{Sun2017}
{Sun} H.,  {Zhang} B.,   {Gao} H.,  2017, \mn@doi [\apj]
  {10.3847/1538-4357/835/1/7}, \href
  {https://ui.adsabs.harvard.edu/abs/2017ApJ...835....7S} {835, 7}

\bibitem[\protect\citeauthoryear{{Sun}, {Li}, {Zhang}, {Zhang}, {Bauer}, {Xue}
  \& {Yuan}}{{Sun} et~al.}{2019}]{Sun2019}
{Sun} H.,  {Li} Y.,  {Zhang} B.-B.,  {Zhang} B.,  {Bauer} F.~E.,  {Xue} Y.,
  {Yuan} W.,  2019, \mn@doi [\apj] {10.3847/1538-4357/ab4bc7}, \href
  {https://ui.adsabs.harvard.edu/abs/2019ApJ...886..129S} {886, 129}

\bibitem[\protect\citeauthoryear{{Suzuki}, {Plucinsky}, {Gaetz}  \&
  {Bamba}}{{Suzuki} et~al.}{2021}]{Suzuki2021}
{Suzuki} H.,  {Plucinsky} P.~P.,  {Gaetz} T.~J.,   {Bamba} A.,  2021, \mn@doi
  [\aap] {10.1051/0004-6361/202141458}, \href
  {https://ui.adsabs.harvard.edu/abs/2021A&A...655A.116S} {655, A116}

\bibitem[\protect\citeauthoryear{{Thompson} et~al.,}{{Thompson}
  et~al.}{2019}]{Thompson2019}
{Thompson} T.~A.,  et~al., 2019, \mn@doi [Science] {10.1126/science.aau4005},
  \href {https://ui.adsabs.harvard.edu/abs/2019Sci...366..637T} {366, 637}

\bibitem[\protect\citeauthoryear{{Weisskopf}, {Tananbaum}, {Van Speybroeck}  \&
  {O'Dell}}{{Weisskopf} et~al.}{2000}]{Weisskopf2000}
{Weisskopf} M.~C.,  {Tananbaum} H.~D.,  {Van Speybroeck} L.~P.,   {O'Dell}
  S.~L.,  2000, in {Truemper} J.~E.,  {Aschenbach} B.,  eds,  Society of
  Photo-Optical Instrumentation Engineers (SPIE) Conference Series Vol. 4012,
  X-Ray Optics, Instruments, and Missions III. pp 2--16 (\mn@eprint {arXiv}
  {astro-ph/0004127}), \mn@doi{10.1117/12.391545}

\bibitem[\protect\citeauthoryear{{Wen} et~al.,}{{Wen} et~al.}{2019}]{Wen2019}
{Wen} J.,  et~al., 2019, \mn@doi [Experimental Astronomy]
  {10.1007/s10686-019-09636-w}, \href
  {https://ui.adsabs.harvard.edu/abs/2019ExA....48...77W} {48, 77}

\bibitem[\protect\citeauthoryear{{Xue} et~al.,}{{Xue} et~al.}{2019}]{Xue2019}
{Xue} Y.~Q.,  et~al., 2019, \mn@doi [\nat] {10.1038/s41586-019-1079-5}, \href
  {https://ui.adsabs.harvard.edu/abs/2019Natur.568..198X} {568, 198}

\bibitem[\protect\citeauthoryear{{Yang}, {Brandt}, {Zhu}, {Bauer}, {Luo}, {Xue}
   \& {Zheng}}{{Yang} et~al.}{2019}]{Yang2019}
{Yang} G.,  {Brandt} W.~N.,  {Zhu} S.~F.,  {Bauer} F.~E.,  {Luo} B.,  {Xue}
  Y.~Q.,   {Zheng} X.~C.,  2019, \mn@doi [\mnras] {10.1093/mnras/stz1605},
  \href {https://ui.adsabs.harvard.edu/abs/2019MNRAS.487.4721Y} {487, 4721}

\bibitem[\protect\citeauthoryear{{Zhang}}{{Zhang}}{2013}]{Zhang2013}
{Zhang} B.,  2013, \mn@doi [\apjl] {10.1088/2041-8205/763/1/L22}, \href
  {https://ui.adsabs.harvard.edu/abs/2013ApJ...763L..22Z} {763, L22}

\bibitem[\protect\citeauthoryear{{Zhang}}{{Zhang}}{2020}]{Zhang2020}
{Zhang} B.,  2020, \mn@doi [\nat] {10.1038/s41586-020-2828-1}, \href
  {https://ui.adsabs.harvard.edu/abs/2020Natur.587...45Z} {587, 45}

\bibitem[\protect\citeauthoryear{{Zhang} et~al.,}{{Zhang}
  et~al.}{2009}]{Zhang2009}
{Zhang} B.,  et~al., 2009, \mn@doi [\apj] {10.1088/0004-637X/703/2/1696}, \href
  {https://ui.adsabs.harvard.edu/abs/2009ApJ...703.1696Z} {703, 1696}

\bibitem[\protect\citeauthoryear{{Zhang} et~al.,}{{Zhang}
  et~al.}{2018}]{Zhang2018}
{Zhang} B.~B.,  et~al., 2018, \mn@doi [Nature Communications]
  {10.1038/s41467-018-02847-3}, \href
  {https://ui.adsabs.harvard.edu/abs/2018NatCo...9..447Z} {9, 447}

\bibitem[\protect\citeauthoryear{{Zhou}, {Chen}, {Li}, {Safi-Harb}, {Mendez},
  {Terada}, {Sun}  \& {Ge}}{{Zhou} et~al.}{2014}]{Zhou2014}
{Zhou} P.,  {Chen} Y.,  {Li} X.-D.,  {Safi-Harb} S.,  {Mendez} M.,  {Terada}
  Y.,  {Sun} W.,   {Ge} M.-Y.,  2014, \mn@doi [\apjl]
  {10.1088/2041-8205/781/1/L16}, \href
  {https://ui.adsabs.harvard.edu/abs/2014ApJ...781L..16Z} {781, L16}

\makeatother
\end{thebibliography}


\bsp	
\label{lastpage}
\end{document}